\documentstyle[PASJadd,psbox]{PASJ95}
\newcommand{\lett}{}

\begin{document}
\title{Discovery of the Slowest X-Ray Pulsar in the SMC, 
AX~J0049.5$-$7323, with ASCA
}
\author{Jun {\sc Yokogawa},
Kensuke {\sc Imanishi},
Masaru {\sc Ueno},
and Katsuji {\sc Koyama}\thanks{CREST, 
Japan Science and Technology Corporation (JST), 
4-1-8 Honmachi, Kawaguchi, Saitama 332-0012.} \\ [12pt]
{\it Department of Physics, Graduate School of Science, Kyoto University, 
Sakyo-ku, Kyoto 606-8502} \\
{\it E-mail(JY): jun@cr.scphys.kyoto-u.ac.jp}}

\abst{The discovery of coherent pulsations with ASCA from an X-ray source, 
AX~J0049.5$-$7323, is reported. 
The barycentric period was determined to be 
$755.5 \pm 0.6$~s, which is the longest among X-ray pulsators in the SMC. 
The X-ray spectrum has been 
found to be unchanged through ASCA observations, 
with a photon index of $\sim 0.8$ and a luminosity of 
$\sim 5 \times 10^{35}$ erg~s$^{-1}$ (0.7--10 keV). 
Archival data of the Einstein and the ROSAT satellites indicate 
that the flux has been 
$\ltsim 10^{-12}$ erg~s$^{-1}$~cm$^{-2}$ 
($\ltsim 5 \times 10^{35}$ erg~s$^{-1}$)
for over 20 years and exhibits a variability with a factor of 
$\gtsim 10$. 
We argue that AX~J0049.5$-$7323 is an X-ray pulsar 
with a Be star companion. 
}

\kword{pulsars: individual (AX~J0049.5$-$7323) --- 
stars: emission-line, Be --- stars: neutron --- X-rays: stars}

\maketitle
\thispagestyle{headings}

%section 1
\section{Introduction}
X-ray binary pulsars (XBPs) are mostly classified 
into OB supergiant binaries and Be X-ray binaries (hereafter Be-XBPs) 
according to their companion stars. 
Traditionally, Be-XBPs 
have been known to exhibit very active outbursts with 
an X-ray luminosity of $\sim 10^{37}$--$10^{38}$ erg~s$^{-1}$, 
which are brighter than the quiescent state 
by a factor of $> 100$--1000 (Stella et al.\ 1986). 
However, 
recently, Mereghetti et al.\ (2000) 
proposed the existence of a subgroup of Be-XBPs 
which show a relatively persistent X-ray luminosity 
with no active outburst. 
Members of this subgroup 
(hereafter, quiescent Be-XBPs), 
such as X~Per and RX~J0146.9$+$6121, 
are characterized by a low-luminosity ($\ltsim 10^{36}$ erg~s$^{-1}$), 
a rather modest flux variability within a factor of $\sim 10$, 
and a long spin period ($\gtsim 300$~s). 
This is probably because 
a magnetized neutron star with a slow rotation 
permits a low accretion rate, i.e., a low luminosity, 
due to a weak centrifugal barrier
at the magnetosphere (Stella et al.\ 1986).

There have been successive discoveries of new XBPs 
in the Small Magellanic Cloud (SMC). 
Most of them have a luminosity 
$\gtsim 10^{36}$ erg~s$^{-1}$ (Yokogawa et al.\ 2000a), 
which probably indicates 
that it has been difficult to detect coherent pulsations from 
a low-luminosity source 
($\ltsim 10^{36}$ erg~s$^{-1}$ or 
$\ltsim 2 \times 10^{-12}$ erg~s$^{-1}$~cm$^{-2}$ 
at the distance of the SMC, 60~kpc; van den Bergh 2000)
because of limited photon statistics. 
Recently, we made an observation on the SMC 
with an exceptionally long exposure time ($\sim 177$~ks), 
and discovered coherent pulsations from 
a low-luminosity source ($\sim 6\times 10^{35}$ erg~s$^{-1}$), 
AX J0051.6$-$7311 (Torii et al.\ 2000; Yokogawa et al.\ 2000b). 
This discovery well demonstrates that 
in the SMC there may be many hidden low-luminosity pulsars, 
including quiescent Be-XBPs, 
and that they will be discovered by observations 
with a very long exposure time. 

In this letter, 
we report on a new X-ray pulsar in the SMC 
having a long period of 755.5~s and a low luminosity 
of $\sim 5 \times 10^{35}$ erg~s$^{-1}$ (Ueno et al.\ 2000), 
discovered during the very long observation. 
The history of the flux variation was also investigated 
with archival data of Einstein Observatory and ROSAT.

%section 2
\section{Observations and Data Reduction}
The position of AX~J0049.5$-$7323 has been covered with 
three ASCA observations. 
The observations spanned during 
50765.179--50766.285 (hereafter obs.\ A1), 
51309.604--51310.682 (obs.\ A2), and 
51645.986--51651.490 (obs.\ A3), in units of Modified Julian Day (MJD). 

ASCA carries four XRTs (X-ray Telescopes, Serlemitsos et al.\ 1995) 
with two GISs (Gas Imaging Spectrometers, Ohashi et al.\ 1996) and 
two SISs (Solid-state Imaging Spectrometers, Burke et al.\ 1994)
on the focal planes. 
Since AX~J0049.5$-$7323 was outside of the field of view (FOV) of the SIS 
in all observations, we do not refer to the SIS any more. 
In each observation, 
the GIS was operated in the normal PH mode 
with a time resolution of 0.0625/0.5~s for a high/medium bit rate. 
We rejected the GIS data obtained in the South Atlantic Anomaly, 
or in low cut-off rigidity regions ($<4$~GV), 
or when the target's elevation angle was low ($< 5^\circ$).
Particle events were removed by a rise-time discrimination method.
After the screening, the total available exposure times 
in obs.\ A1, A2, and A3 were
$\sim43$~ks, $\sim 41$~ks, 
and $\sim177$~ks, respectively.

%section 3
\section{Results}
%section 3.1
\subsection{Source Identification}
In obs.\ A3, a faint source was detected at $\sim 15'$ from 
the center of FOV. 
Its position was determined 
with the method described by Ueda et al.\ (1999) 
to be 
(00$^{\rm h}$49$^{\rm m}$33$^{\rm s}$, $-$73$^\circ$23$'$23$''$) 
for equinox 2000.0; 
we thus designate this source as AX~J0049.5$-$7323. 
Note that this source was originally named AX~J0049.4$-$7323 
in Ueno et al.\ (2000). 
The radius of the error circle would be $\sim 1'\hspace{-2pt}.5$, 
because of the large off-axis angle. 
This source was also detected in obs.\ A1 and A2
at $\sim 16'$ and $\sim 22'$ from the center of FOV, respectively. 
Here and in the following analyses, 
we used GIS 2 $+$ GIS 3 data for obs.\ A1 and A3, 
while only GIS 2 was used for obs.\ A2, because 
AX~J0049.5$-$7323 was located near to the calibration source of GIS 3.

We investigated catalogs of X-ray sources detected with the Einstein 
and the ROSAT satellites (Wang, Wu 1992; Haberl et al.\ 2000). 
Only one ROSAT source, No.\ 468 in Haberl et al.\ (2000), 
was found in the ASCA error circle. 
%No.468 = 00 49 43.8 -73 23 02

%section 3.2
\subsection{Timing Analyses}
In each observation, 
source photons were collected from a circle 
with a radius of $\sim 3'$ 
centered on AX~J0049.5$-$7323. 
We made a barycentric correction on the photon arrival times 
and performed an FFT (Fast Fourier Transformation) 
on the event lists in each observation. 
A significant peak was detected at a frequency of $\sim 0.0013$~Hz
only in the power spectrum of obs.\ A3 (figure 1). 
A maximum power of 49.12 was obtained from events in 
the energy band of 1.2--5.5~keV. 
The probability to detect such a large power in any frequency 
from random events is estimated to be very low, $\sim 1 \times 10^{-5}$; 
we thus consider that pulse detection is significant. 
We then performed an epoch folding search and determined 
the barycentric period to be $P = 755.5 \pm 0.6$~s, 
which is the longest among the X-ray pulsators known in the SMC. 
We did not find any evidence for pulsations from obs.\ A1 and A2, 
either by an FFT analysis or by an epoch folding search.

Figure 2 shows the pulse profiles during obs.\ A3
in the energy bands of 
1.2--2.0~keV, 2.0--5.5~keV, and 5.5--10.0~keV. 
The pulse shape is rather spiky in 1.2--2.0~keV, while 
it is nearly sinusoidal in 2.0--5.5~keV. 
It should also be noticed that 
higher energy photons ($\gtsim 5.5$~keV) 
do not exhibit definite pulsations. 
The pulsed fraction, defined as (pulsed flux)/(total flux) without 
background, is $\sim 45$\% in 2.0--5.5~keV.

%Place fig. 1 here.
%Place fig. 2 here.
\begin{figure}
\hspace*{8mm}\psbox[xsize=0.4\textwidth]{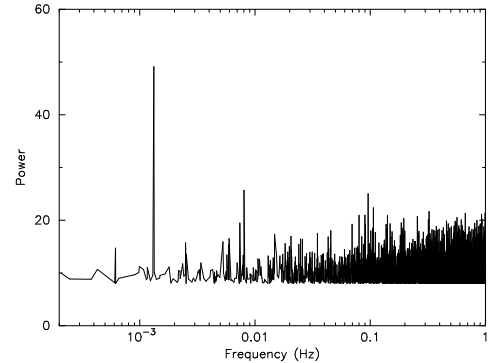}
\caption{Power spectrum in obs.\ A3, in which 
power is normalized so that the random-fluctuation level corresponds to 2. 
Data points with a power less than 8 are omitted. 
An evident peak is detected at $\sim 0.0013$~Hz.}
\end{figure}

\begin{figure}
\hspace*{8mm}\psbox[xsize=0.4\textwidth]{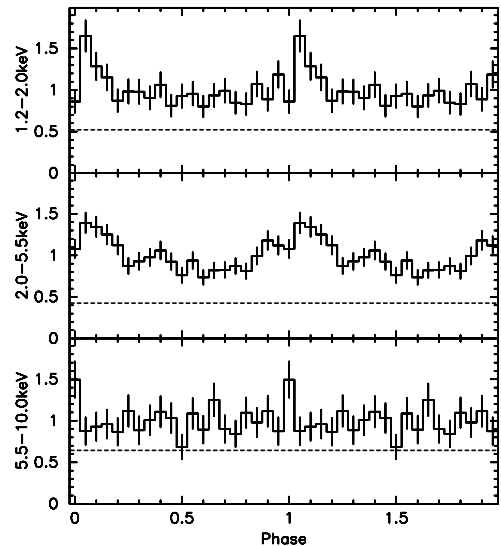}
\caption{Pulse profiles in obs.\ A3. 
Upper, middle, and lower panels are for 
the 1.2--2.0~keV, 2.0--5.5~keV, and 
5.5--10.0~keV bands, respectively.
The vertical axes indicate the normalized count rates in each energy band. 
The background levels are indicated by the broken lines.}
\end{figure}

We also searched for an aperiodic intensity variation 
during each observation, but neither bursts nor large a flux variation 
was detected.

%section 3.3
\subsection{Spectral Analyses}
The source spectrum in each observation 
was extracted from the same region used in 
the timing analyses, 
while the background spectrum was 
from an off-source area near to the source.

At first, 
we separately fitted each spectrum 
to a simple power-law model with the interstellar absorption. 
The derived parameters (photon index $\Gamma$ and column density $N_{\rm H}$) 
were found to be consistent throughout all observations within 
the statistical errors. 
Therefore, 
to strictly constrain the parameters, 
we assumed that $\Gamma$ and $N_{\rm H}$ were identical 
in all observations, and simultaneously fitted all of the spectra
to the same model. 
The best-fit model was acceptable at the 90\% confidence level 
with the following parameters: 
$\Gamma = 0.78$ (0.66--1.00) and $N_{\rm H} = 0 (<3) \times 10^{21}$ 
cm$^{-2}$, with a reduced $\chi^2$ of 1.16 for 102 degrees of freedom
(values in parentheses indicate 90\% confidence limits). 
The fluxes (0.7--10.0~keV) were derived to be 
1.1, 1.5, and 1.7 
for obs.\ A1, A2, and A3, respectively, 
in units of $10^{-12}$ erg~s$^{-1}$~cm$^{-2}$. 
Figure 3 shows the phase-averaged GIS spectrum 
and the best-fit model in obs.\ A3, 
which had the best statistics.

%Place fig. 3 here.
\begin{figure}
\hspace*{8mm}\psbox[xsize=0.4\textwidth]{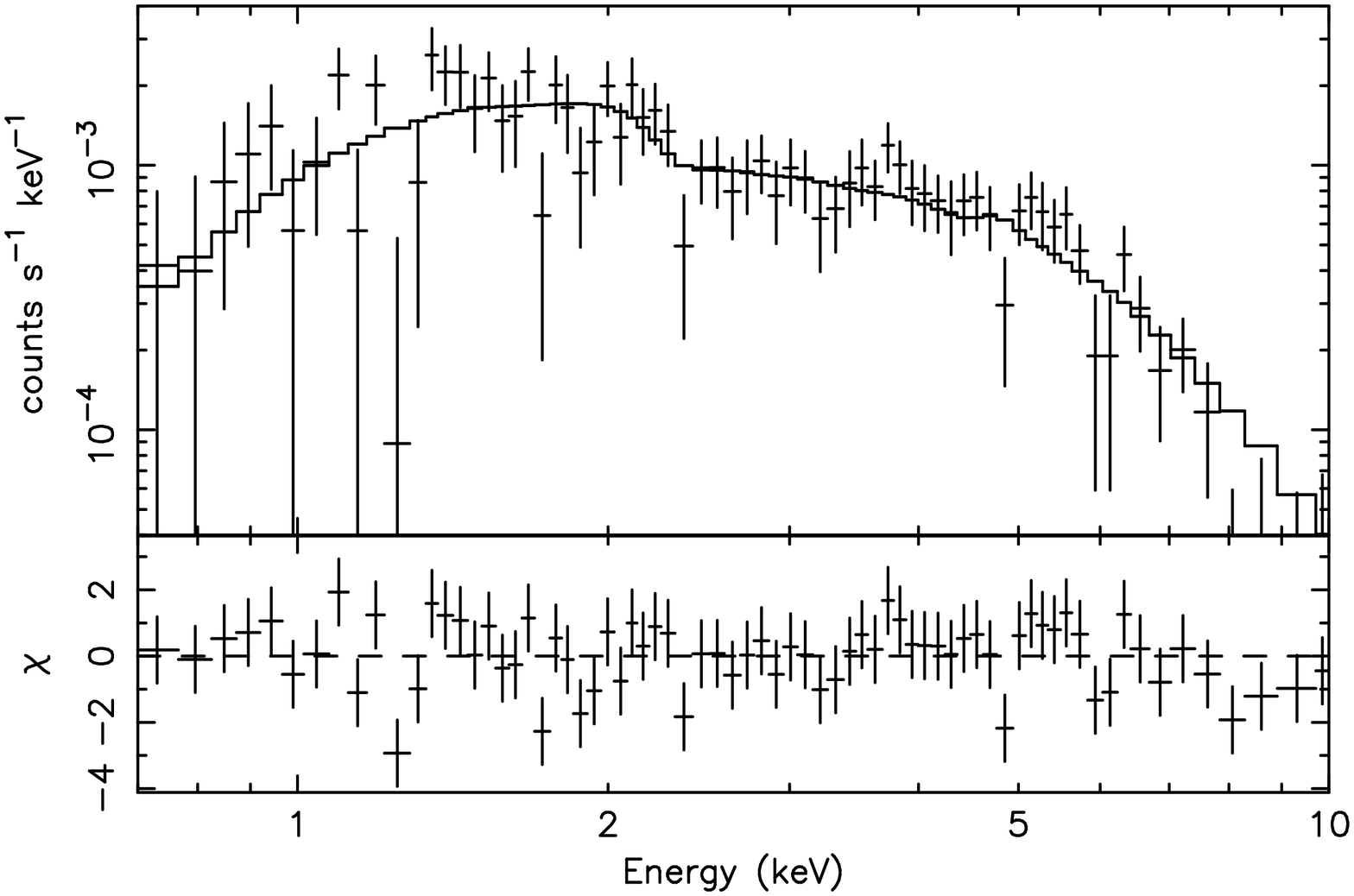}
\caption{Background-subtracted and phase-averaged spectrum in obs.\ A3, 
obtained with GIS 2+3. 
The crosses and the solid line indicate 
the data points and the best-fit model, respectively. 
}
\end{figure}

For obs.\ A3, 
we extracted phase-resolved spectra from 
phase 0--0.2, which represents the ``spike'' 
seen in 1.2--2.0~keV, and phase 0.2--1 
(see figure 2), 
and fitted them with the power-law model. 
The best-fit parameters may indicate a slight change in the spectrum: 
$\Gamma = 1.4$ (0.9--2.1) and 
$N_{\rm H} = 9 $ (2--$20) \times 10^{21}$ cm$^{-2}$ in phase 0--0.2, 
while 
$\Gamma = 0.8$ (0.6--0.9) and 
$N_{\rm H} = 0 (<2) \times 10^{21}$ cm$^{-2}$ in phase 0.2--1.

%section 4
\section{Discussion}
AX~J0049.5$-$7323 has been covered by 
3 and 16 observations with Einstein and ROSAT, respectively. 
We used these archival data to investigate any long-term flux variation 
of this source. 
We first derived the count rate from AX~J0049.5$-$7323 
after background subtraction 
and a vignetting correction 
in each observation. 
We then converted the count rate to the X-ray flux
with the {\tt PIMMS} software, 
assuming that $\Gamma$ and $N_{\rm H}$ 
are the same as those derived from ASCA observations 
($\Gamma = 0.78$ and $N_{\rm H} = 0$ cm$^{-2}$). 
The results are shown in figure 4; 
we found that AX~J0049.5$-$7323 
has been detected with a flux of 
$\ltsim 10^{-12}$ erg~s$^{-1}$~cm$^{-2}$, 
and occasionally it went down below a detection upper limit of 
$\sim 10^{-13}$ erg~s$^{-1}$~cm$^{-2}$. 
We can point out 
a flux variability with a factor of $\gtsim 20$ 
from figure 4. 
However, we should note that 
the assumption of $N_{\rm H}$ has some impact on 
the Einstein/ROSAT fluxes in this procedure, 
because the energy ranges covered by Einstein and ROSAT 
extend well below that of ASCA. 
To estimate the effect of $N_{\rm H}$, 
we assumed $N_{\rm H} = 3 \times 10^{21}$ cm$^{-2}$ 
(upper limit derived in the spectral analysis) 
and derived the Einstein/ROSAT fluxes with {\tt PIMMS} again. 
We found that the fluxes increase by a factor of 
$\sim 1.5$--2.5. 
Taking these facts into account, 
we conclude that 
AX~J0049.5$-$7323 has been faint 
($\ltsim 10^{-12}$ erg~s$^{-1}$~cm$^{-2}$) for $\sim 20$~yr 
and has experienced a flux variation 
of a factor of at least $\sim 10$.

%Place fig. 4 here.
\begin{figure}
\hspace*{8mm}\psbox[xsize=0.4\textwidth]{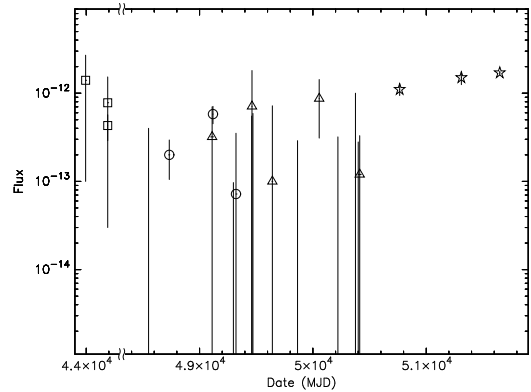}
\caption{Flux history of AX~J0049.5$-$7323 
after vignetting correction
(in units of erg~s$^{-1}$~cm$^{-2}$, 0.7--10.0~keV). 
Data were obtained with 
Einstein/IPC (squares), 
ROSAT/PSPC (circles),
ROSAT/HRI (triangles), and 
ASCA/GIS (stars). 
Error bars indicate 1-$\sigma$ statistical errors. 
The data points for which only error bars are given 
indicate no detection. 
}
\end{figure}

We detected coherent pulsations only from obs.\ A3. 
It would be attributed to the much better statistics in obs.\ A3, 
which has a very long exposure time ($\sim 177$~ks), 
or it may imply a smaller pulsed fraction in obs.\ A1 and A2. 
To examine the former possibility, 
we divided obs.\ A3 into four segments, 
each having an exposure time of $\sim 44$~ks, 
and performed the FFT and the epoch folding search 
on each segment. 
We detected no significant pulsations from any segment. 
Therefore, we conclude that a very long observation was essential 
for pulse detection from this source. 
The same condition applies to another new pulsar, AX~J0051.6$-$7311 
(Yokogawa et al.\ 2000b).

We discuss the nature of AX~J0049.5$-$7323. 
The pulse period of 755.5~s is common with 
both XBPs and white dwarf (WD) binaries. 
However, the X-ray spectra showing no emission line 
and its luminosity of 
$\sim 5 \times 10^{35}$ erg~s$^{-1}$ 
($\sim 1 \times 10^{-12}$ erg~s$^{-1}$~cm$^{-2}$ in flux)
are rather typical for XBPs (e.g., Stella et al.\ 1986), 
and against the possibility of a WD binary, 
which has a spectrum dominated by strong emission lines 
from ionized atoms 
and a luminosity of $\ltsim 10^{33}$ erg~s$^{-1}$
(see Ezuka, Ishida 1999). 
In addition, the ROSAT counterpart, 
No.\ 468 in Haberl et al.\ (2000), 
is proposed to be a Be/X-ray binary 
(Haberl, Sasaki 2000), 
because an emission line object is found in 
the error circle ($14''\hspace{-3pt}.9$ radius) determined with ROSAT. 
Therefore, we would conclude that 
AX~J0049.5$-$7323 is a Be-XBP with the longest spin period 
in the SMC.

We point out that AX~J0049.5$-$7323 shares 
some characteristics with quiescent Be-XBPs in our Galaxy: 
the long pulse period, 
no large outburst, 
and small and relatively persistent luminosity 
(although the lower limit could not be determined). 
However, 
the strong energy dependence of the pulse shape 
(figure 2) in AX~J0049.5$-$7323 is rather peculiar, 
when compared with five quiescent Be-XBPs 
(X~Per, RX~J0146.9$+$6121, RX~J0440.9$+$4431, RX~J1037.5$-$564, 
and 1SAX~J0103.2$-$7209 ---  
White et al.\ 1983; Mereghetti et al.\ 2000; Reig, Roche 1999; 
Israel et al.\ 2000). 
Whether AX~J0049.5$-$7323 is unusual or not 
will be examined by 
studying the energy-resolved pulse shapes of other quiescent Be-XBPs. 
\par
\vspace{1pc}\par
We thank Dr.\ Torii and the referee, Dr.\ Ishida, 
for their useful and constructive suggestions. 
The Einstein and ROSAT data were obtained through the High Energy
Astrophysics Science Archive Research Center Online Service,
provided by the NASA/Goddard Space Flight Center. 
J.Y.\ is 
supported by JSPS Research Fellowship for Young Scientists.

%\clearpage
\section*{References} 
%\small %½ÐÈÇ»þ
\re
 Burke B.E., Mountain R.W., Daniels P.J., Dolat V.S., 
 Cooper M.J.\ 1994, IEEE Trans.\ Nucl.\ Sci.\ 41, 375
\re
 Ezuka H., Ishida M.\ 1999, ApJS 120, 277
\re
 Haberl F., Filipovi\'{c} M.D., Pietsch W., Kahabka P.\ 
 2000, A\&AS 142, 41
\re
 Haberl F., Sasaki M.\ 2000, A\&A 359, 573
\re
 Israel G.L., Campana S., Covino S., Dal Fiume D., Gaetz T.J., Mereghetti S., 
 Oosterbroek T., Orlandini M.\ et al.\ 2000, ApJ 531, L131
%, Parmar A.N., Ricci D., Stella L.
\re
 Mereghetti S., Tiengo A., Israel G.L., Stella L.\ 2000, A\&A 354, 567
\re
 Ohashi T., Ebisawa K., Fukazawa Y., Hiyoshi K., 
 Horii M., Ikebe Y., Ikeda H., Inoue H.\ 
 et al.\ 1996, PASJ 48, 157
\re
 Reig P., Roche P.\ 1999, MNRAS 306, 100
\re
 Serlemitsos P.J., Jalota L., Soong Y., Kunieda H., Tawara Y., 
 Tsusaka Y., Suzuki H., Sakima Y.\ 
 et al.\ 1995, PASJ  47, 105
\re
 Stella L., White N.E., Rosner R.\ 1986, ApJ 308, 669
\re
 Torii K., Yokogawa J., Imanishi K., Koyama K.\ 2000, IAU Circ.\ 7428
\re
 Ueda Y., Inoue H., Ogawara Y., Fujimoto R., Yamaoka K., Kii T., 
 Gotthelf E.V.\ 1999, ISAS research Note 688
\re
 Ueno M., Yokogawa J., Imanishi K., Koyama K.\ 2000, IAU Circ.\ 7442
\re
 van den Bergh, S.\ 2000, PASP 112, 529
\re
 Wang Q., Wu X.\ 1992, ApJS 78, 391
\re
 White N.E., Swank J.H., Holt S.S.\ 1983, ApJ 270, 711
\re
 Yokogawa J., Imanishi K., Tsujimoto M., Nishiuchi M., Koyama K., 
 Nagase F., Corbet R.H.D.\ 2000a, ApJS 128, 491
\re
 Yokogawa J., Torii K., Imanishi K., Koyama K.\  2000b, 
 PASJ 52, L37

%\clearpage
%when submitting...
%\centerline{Figure Captions}
%\bigskip
%
%
%

\end{document}